# Cultivating Metal Whiskers by Surface Plasmon Polariton Excitation

Vamsi Borra[1], Daniel G. Georgiev[1] and Victor G. Karpov[2]

[1] Department of EECS, University of Toledo, Toledo, OH 43606-3390, U.S.A.

[2] Department of Physics and Astronomy, University of Toledo, Toledo, OH 43606-3390, U.S.A.

**ABSTRACT**

This work presents a preliminary experimental study on the possibility to initiate growth of whiskers on the surfaces of some technologically important metals utilizing the enhanced electric field of surface plasmon polaritons (SPPs). The results provide evidence that a relatively high concentration of what appear to be whisker nuclei form in the region where SPPs were excited, whereas no such changes are observed on the untreated surface.

**INTRODUCTION**

Metals that are extensively used in electronics manufacturing, such as tin (Sn), zinc (Zn), and related alloys, often show electrically conductive hair-like crystalline structures on their surfaces, referred to as whiskers. Whiskers can lead to current leakage and short circuits in sensitive electronic equipment causing significant losses and, in some cases, catastrophic failures in the automotive, airspace and other industries[2,3]. The mechanism responsible for whisker formation and growth remains mysterious after decades of research. Some theories attribute metal whisker growth to stress relieving phenomena, but those never lead to any quantitative estimate of whisker parameters. A recent electrostatic theory[3] provides a quantitative estimates of the whisker nucleation along with their growth rates and length distributions, which are consistent with many observations on whiskers growth and morphology. It proposes that the imperfections on metal surfaces can form small patches of net positive or negative electric charge leading to the formation of the anomalous electric field which governs the whisker development in those areas. An external electric field, which can be either constant (DC) or varying with time (AC) at high frequencies (including optical), can then further metal whiskers growth by means of lowering the free energy of the system.

In this work, which is still in progress, we report on the possibility of initiating and controlling the growth of whiskers by taking advantage of the enhanced AC electric field at the metal surface on which SPPs are generated. The Otto and Kretschmann prism-based methods are well established and studied attenuated total reflection geometries for indirect optical excitation of SPPs on the surface of metals and metal films. Generating SPPs on low-loss metal films such as Ag and Au is generally straightforward. However, generating SPPs on the surfaces of metals with large damping losses, such as Sn and Zn, is challenging as the energy of the incident field is largely absorbed in the process of coupling to the SPPs[4], [5]. In the present study, optically smooth Sn and Zn thin, films fabricated by vacuum evaporation, are subjected to SPPs excitation to begin the whisker growth. We used optical and electron microscopy techniques to confirm the nucleation.

# EXPERIMENTAL DETAILS

Because of conservation laws requirements, direct coupling between the photon and SPPs is unattainable. Prism coupling techniques or near field probe based ones can be used to address this problem, i.e., providing momentum matching[4], [6] [7], [8]. Otto and Kretschmann geometries have been extensively used for generating SPPs on the metal surfaces as reported in most of the available literature. In our experiment, Otto configuration suits best because it provides a reasonable SPPs propagation range and also offers the opportunity of using relatively thick metal films or bulk metal samples, even when accounting for the intrinsic experimental difficulty of maintaining a required micron-scale gap ($t_d$ in Fig.1). Our experimental setup is schematically represented in figure1. Here, $\varepsilon_p$, $\varepsilon_d$ and $\varepsilon_m$ are the dielectric constants of prism, dielectric material and metal film respectively. $t_d$ and $t_m$ are thickness of dielectric material and metal film.

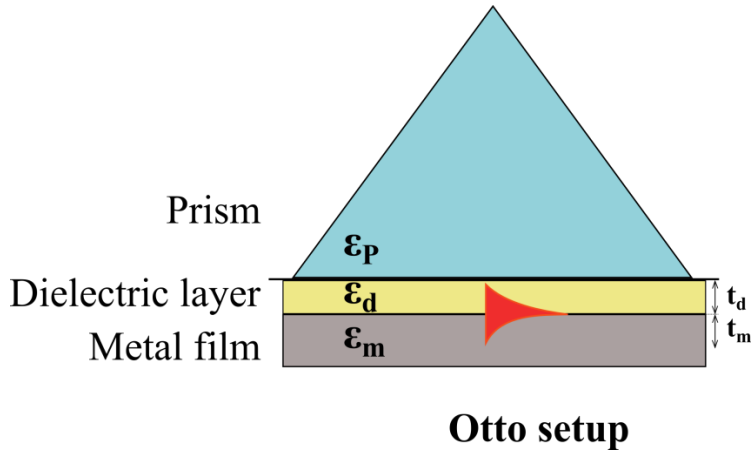

**Figure 1. Experimental setup.**

The fabrication of optically smooth Sn films by vacuum based methods turned out to be somewhat challenging but we developed a method for achieving it[9]. Thin films are mounted on to the plano side of a UV grade fused silica plano-convex cylindrical lens serving as the Otto prism (from Thorlabs, n=1.460 at 588nm). Air plays the role of a dielectric layer between the metal and the prism for our case. We employed a 100mW CW semiconductor laser system ($\lambda$=662nm, model OBIS 660 by Coherent) to excite the SPPs. The reflected power from the metal surface with respect to various incident angles is recorded using a silicon photodiode based power probe (from Coherent) to verify the generation of SPPs on the metal film. The SPPs excitation experiment was first performed on noble metal films, such as Ag (small damping loss), in order to verify our ability to form the SPPs with our setup and then later reproduced the same on the Sn metal which has high damping loss. The exposed surfaces were examined by optical microscopy and scanning electron microscopy (SEM) subsequently for identifying the nucleation site of the whiskers.

# DISCUSSION

If the polarized laser beam with electric field in TM mode is incident on the metal film, which is mounted on the prism, and its angle of incidence ($\Theta_i$) is slightly greater than the critical angle for total reflection, an evanescent wave is formed on the metal surface. The evanescent wave can excite the natural modes of oscillations of the electrons at the metal surfaces. These collective oscillations form SPPs that propagate parallel to the surface but decay very quickly in the direction normal to the surface. Their existence can be indirectly probed by recording reflectance as a function of the angle of incidence. The deep minimum observed beyond the critical angle in the recorded reflectance curve authenticates their presence. Some literature suggests that the angle of incidence at which the reflectivity exhibits a deep minimum is not the same as the angle ($\Theta_p$) of incidence at which SPPs have maximum intensity. The difference in $\Theta_i$ and $\Theta_p$ ranges in $\pm 0.3°$ for noble metals (Ag, Au, etc.) due to their small damping losses;, unfortunately it may vary between $25°$ - $35°$ for metals with large damping losses (Fe, Sn, Zn, etc.) [4], [5].

## Theoretical calculations

The field enhancement between the prism and the metal surface can be calculated with the Fresnel reflection and transmission coefficients. The Fresnel reflection $R_{123}$ and transmission $T_{123}$ coefficient equations for our experimental setup are derived using equations 2 and 3 in Ref. [10].

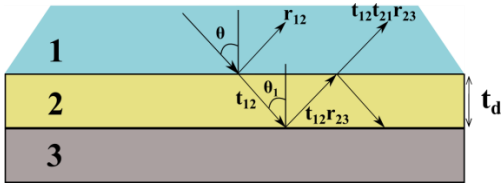

**Figure 2. Three-layered dielectric structure.**

In figure 2, $r_{12}$ and $t_{12}$ are the Fresnel coefficients between the prism (layer-1) and air (layer-2). Similarly, $r_{23}$ and $t_{23}$ are the Fresnel coefficients between the air (layer-2) and metal (layer-3). Here $t_d$ is the distance that separates the prism and the metal surface. A computer program is formulated using the above mentioned equations to find the optimized value of $t_d$ such that the magnitude of the reflected wave is at its minimum. The program was designed to solve the above equations with a range of $t_d$ values by varying the incident angle of the laser beam. These simulations have been performed for Ag and Sn metal surfaces separately.

## Simulation and Experiment on Ag metal film

The dielectric function of the metallic thin film is denoted by $\varepsilon = \varepsilon_1 + i\varepsilon_2$. The imaginary part ($\varepsilon_2$) of the dielectric function plays a crucial role in the optical absorption in the metal. At our laser wavelength ($\lambda$=662nm), the dielectric function of Ag is taken from this literature[11]. The refractive index for prism and air are considered as $n_p$=1.460 (from manufacturer) and $n_{air}$=1 respectively. Upon solving the Fresnel equations with our program, $t_d = 0.8\mu m$ and the deep minimum in the reflectivity when incidence angle is $45.09°$ are obtained, shown in figure 3(a). In the experiment the deep minimum occurred at incidence angle of 48.67°, shown in figure 3(b). This slight difference can be attributed to the inaccuracy in mounting the prism onto the rotating stage. Overall, this observation validates our experimental setup and the proof that we are generating SPPs on Ag metal film.

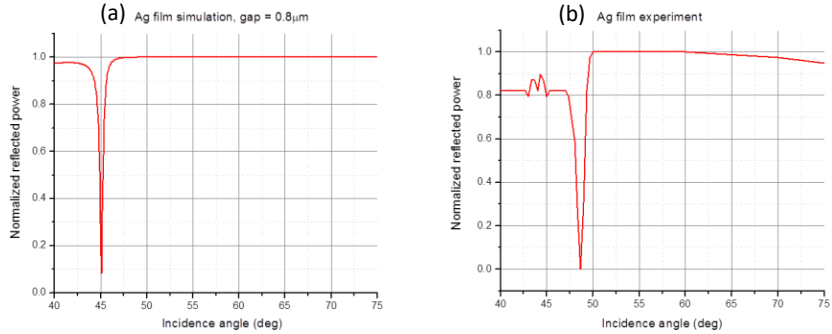

**Figure 3. Reflectance vs. angle of incidence for Ag films (a) Simulation (b) Experiment.**

### Simulation and Experiment on Sn metal film

Having successfully generated SPPs on the low damping Ag films, we performed a similar experiment on the high damping Sn metal films. At λ=662nm, the dielectric function of Sn is taken from this literature[12]. As expected, the SPP dip is shifted significantly (shown in figure 4) due to a larger damping. From the experimental plots we conclude that the SPPs occurred when incidence angle is 68°. This shift can also be observed on other metals (Fe, Zn etc.) that have significantly large imaginary part ($\varepsilon_2$) in the dielectric function[4], [5]. For our case the Sn's $\varepsilon_2$ is around 13 times the value of Ag.

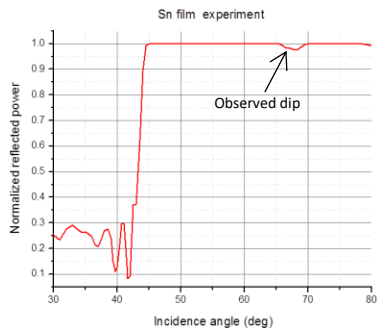

**Figure 4. Experimental Reflectance vs. angle of incidence for Sn films.**

### Scanning electron microscope scans

Having identified the deep minimum experimentally, we left the laser at that incidence angle (68°) for 5 hours at 100mW incident power. The figure 5 and 6 are the scanning electron microscope (SEM) images of the sample outside the irradiated spot and on the spots respectively. The scanned area in figure 5 doesn't show any sign of surface features formation away from the irradiated spot. To the contrary, the irradiated spot in the figure 6(a) and 6(b) show the clear sign of localized features that we attribute to whisker nuclei. The observed densities of such nuclei are quite high at the irradiated spot. A closer look (shown in figure 6(b)) at those nuclei reveals whisker like structure growing vertical to the sample surface. The direction of whisker growth aligns well with the SPPs field direction.

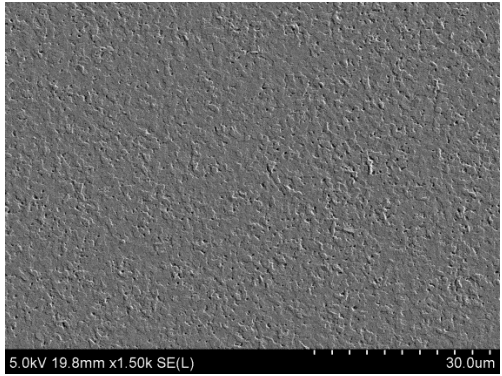

**Figure 5. No sign of nuclei.**

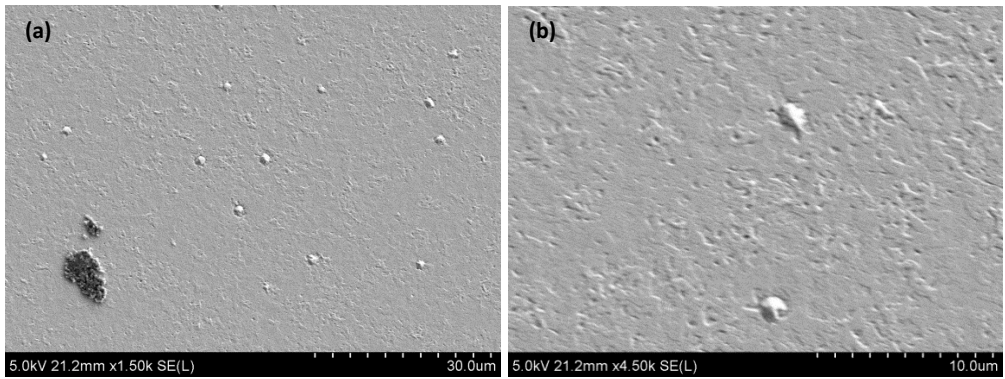

**Figure 6. SEM scans of the laser irradiated samples.**

## CONCLUSIONS

This experimental observation has important implications: (1) it can experimentally verify some aspects of the electrostatic theory of metal whiskers using SPPs and (2) it paves a way to the controllable formation of whiskers. The latter would allow the development of accelerated failure testing methods of electronic components, which is currently not possible due to the random and unpredictable nature of whiskers formation. We also anticipate that these nuclei can be grown into a fully grown whisker with further applied electric field; this part of our work is still in progress.

## ACKNOWLEDGMENTS

The authors would like to thank the University of Toledo's EECS department, and the University of Toledo Graduate Student Association (GSA) for providing the financial support to

**REFERENCES**

[1] NASA, "NASA Goddard Space Flight Center tin whisker homepage." [Online]. Available: http://nepp.nasa.gov/whisker/failures/index.htm. [Accessed: 11-Apr-2015].

[2] J. Brusse, "Metal Whisker Discussion (Tin and Zinc Whiskers)," 2004. [Online]. Available: http://www.calce.umd.edu/lead-free/other/BRUSSE_ACI.pdf. [Accessed: 11-Apr-2015].

[3] V. G. Karpov, "Electrostatic Theory of Metal Whiskers," *Phys. Rev. Appl.*, vol. 1, no. 4, p. 044001, May 2014.

[4] E. F. Y. Kou and T. Tamir, "Incidence angles for optimized ATR excitation of surface plasmons," *Appl. Opt.*, vol. 27, no. 19, pp. 4098–4103, 1988.

[5] R. D. Olney and R. J. Romagnoli, "Optical effects of surface plasma waves with damping in metallic thin films.," *Appl. Opt.*, vol. 26, no. 11, pp. 2279–82, 1987.

[6] W. L. Barnes, A. Dereux, and T. W. Ebbesen, "Surface plasmon subwavelength optics," *Nature*, vol. 424, no. 6950, pp. 824–830, 2003.

[7] D. Sarid, "Long-Range Surface-Plasma Waves on Very Thin Metal Films," *Phys. Rev. Lett.*, vol. 47, no. 26, pp. 1927–1930, Dec. 1981.

[8] F. Y. Kou and T. Tamir, "Range extension of surface plasmons by dielectric layers," *Opt. Lett.*, vol. 12, no. 5, pp. 367–369, May 1987.

[9] V. Borra, D. Georgiev, and V. Karpov, "Fabrication of optically smooth Sn thin films ( Manuscript in preparation)," 2015.

[10] T. Tsang, T. Srinivasanrao, and J. Fischer, "Surface-Plasmon Field-Enhanced Multiphoton Photoelectric-Emission From Metal-Films," *Phys. Rev. B*, vol. 43, no. 11, pp. 8870–8878, 1991.

[11] A. D. Rakić, A. B. Djurišić, J. M. Elazar, and M. L. Majewski, "Optical properties of metallic films for vertical-cavity optoelectronic devices," *Appl. Opt.*, vol. 37, no. 22, pp. 5271–5283, Aug. 1998.

[12] E. T. A. R. A. MacRae, "Optical Properties of Vacuum-Evaporated White Tin*," *Phys. Rev.*, vol. 162, no. 3, pp. 615–620, 1967.